\documentclass[12pt]{iopart}
\usepackage{graphicx}
\usepackage{citesort}
\usepackage{color,soul}
\usepackage{amssymb}

\begin{document}

\title[]
  {Surface effects on ionic Coulomb blockade in nanometer-size pores}

\author{Hiroya Tanaka}
\address{Toyota Central Research \& Development Labs. Inc., Nagakute, Aichi 480 1192, Japan}
\ead{tanak@mosk.tytlabs.co.jp}

\author{Hideo Iizuka}
\address{Toyota Central Research \& Development Labs. Inc., Nagakute, Aichi 480 1192, Japan}

\author{Yuriy V. Pershin}
\address{Department of Physics and Astronomy, University of South Carolina, Columbia, SC 29208, USA}

\author{Massimiliano Di Ventra}
\address{Department of Physics, University of California, San Diego, La Jolla, CA 92093, USA}

\vspace{10pt}
\begin{indented}
\item[]April 2017
\end{indented}

\begin{abstract}
Ionic Coulomb blockade in nanopores is a phenomenon that shares some similarities but also differences with its electronic counterpart. Here, we investigate
extensively this phenomenon using all-atom molecular dynamics of ionic transport through nanopores of about one nanometer in diameter and up to several nanometers in length. Our goal is to better understand the role of
atomic roughness and structure of the pore walls in the ionic Coulomb blockade. Our numerical results reveal the following general trends. First,
the nanopore selectivity changes with its diameter, and the nanopore position in the membrane influences the current strength.
Second, the ionic transport through the nanopore takes place in a hopping-like fashion
over a set of discretized states caused by local electric fields due to membrane atoms. In some cases, this creates a slow-varying ``crystal-like'' structure of ions inside the nanopore. Third,  while at a given voltage, the resistance of the nanopore depends on its length, the slope of this dependence appears to be independent
of the molarity of ions. An effective kinetic model that captures the ionic Coulomb blockade
behavior observed in MD simulations is formulated.
\end{abstract}

%
\vspace{2pc}
\noindent{\it Keywords}: Nanopore, ionic transport, Coulomb blockade
%
%
%
%

\section{Introduction}

The phenomenon of electronic Coulomb blockade, first introduced in 1950s~\cite{Gorter51}, still attracts significant attention. Examples of systems and devices exhibiting the Coulomb blockade include quantum dots~\cite{Beenakker91a}, tunnel junctions~\cite{Averin86a}, molecular systems~\cite{Park02a}, graphene nanostructures~\cite{Stampfer08a}, spintronics devices~\cite{pershin07a,pershin08a}, etc. Moreover, the Coulomb blockade is not unique to electrons being observed, e.g., for both electrons and holes in heterojunctions~\cite{Imamoglu94a}.

The field of {\it ionic} Coulomb blockade is instead relatively new~\cite{R16} with a small number of studies reported in the literature so far. It was predicted theoretically by Krems and Di Ventra while studying ionic transport through a V-shaped nanopore~\cite{R16}. Later, this work was extended to biological ion channels~\cite{Kaufman15a}. Recently, ionic Coulomb blockade has been demonstrated experimentally using a nanopore in a MoS$_2$ membrane~\cite{R17}. Theoretically, the traditional capacitive model \cite{R16,Kaufman15a,R17} and a kinetic approach~\cite{R16} were used in studies of ionic Coulomb blockade.


While the ionic and electronic Coulomb blockades are related phenomena, their experimental manifestations are not exactly the same. For example, {\it electronic} Coulomb blockade in tunneling junctions separating a central island from the electrodes is usually associated with gaps between current (conductance) peaks on the current (conductance) versus gate voltage plots~\cite{C5NR08954A}. The gaps appear at gate voltages such that an extra charge cannot tunnel through the central island if not enough energy is provided to it~\cite{R19}. Moreover, the electronic Coulomb blockade is observed at low temperatures (e.g., of the order of 1 K)~\cite{R19}. Instead, in the case of {\it ionic} Coulomb blockade, the measurements can be performed at room temperature with ions dissolved in a solvent (e.g., water),
and it is realized at the classical level with no tunneling involved. A non-linear dependence of the current on ionic molarity with saturation is considered as a signature of this phenomenon~\cite{R16,R17}. Typically, the ionic Coulomb blockade is associated with trapping/accumulation of ions inside the nanopore impeding the flow of other same-type ions due to Coulomb repulsion~\cite{R16}.

The strength of the ionic Coulomb blockade -- the Coulomb ``gap" energy (the extra energy needed to add another ion to the nanopore) -- can be estimated by~\cite{Kaufman15a}
\begin{equation}
U=\frac{z^2e^2}{2C}=\frac{1}{4\pi\varepsilon_0}\frac{z^2e^2L}{2\varepsilon_w r^2},
\label{Coul_gap}
\end{equation}
where $e$ is the elementary charge, $\varepsilon_0$ is the permittivity of free space,
 $\varepsilon_w$ is the relative permittivity of water, $C$ is the geometry-dependent
self-capacitance of the nanopore given by $4\pi\varepsilon_0 \varepsilon_w r^2/L$, $z$ is the ion
valence number, $L$ is the length of the nanopore and $r$ is its radius. From general considerations, the Coulomb blockade
regime is expected when the sum of the thermal energy per particle plus the extra ``drift'' energy supplied by, e.g., an electric field,  is smaller than $U$~\cite{R16}. Assuming the drift energy to be much smaller than the thermal energy, we can estimate the relevant nanopore sizes to observe this phenomenon. By taking $L\sim r$ and using
$z=1$, $\varepsilon_w=80$,  $T=293$ K, one finds $r \lesssim r^*=0.36$ nm.

However, these simple estimates do not fully account for the actual atomic structure of realistic nanopores. For instance, it is not
at all clear what the role of surface corrugation is, where by surface corrugation we mean the one pertaining to the {\it internal} walls of the nanopore. Atomic corrugation may lead to local fields that trap ions locally in space and for some intervals of time. These effects may indeed reveal other aspects of the phenomenon.

The goal of this study is to develop a better understanding of the ionic Coulomb blockade in realistic nanometer-diameter pores. For this purpose, we performed extensive all-atom molecular dynamics (MD) simulations of ionic transport by varying multiple parameters such as the membrane thickness, nanopore position and diameter, ionic concentration, applied bias, etc. These simulations have allowed us to develop a comprehensive picture of ionic transport in realistic nanopores.

We find that
the nanopore selectivity changes with its diameter, and the nanopore position in the membrane influences the current strength.
In addition, we find that the ionic transport through the nanopore takes place in a hopping-like fashion
over a set of discretized states caused by local electric fields due to membrane atoms. This may cause a slow-varying ``crystal-like'' structure of ions inside the nanopore. At a given voltage, the resistance of the nanopore depends on its length, while the slope of this dependence appears to be independent
of the molarity of ions. We have finally developed an effective kinetic model that captures the ionic Coulomb blockade
behavior observed in MD simulations.

The paper is organized as follows. In the next section we introduce the system under study and provide details of MD simulations. Then, we present simulation results for representative nanopores. First of all, we consider the role of nanopore diameter and location in nanopore selectivity and capacity to carry the current. Second, we focus on details of  ionic transport in nanopores. Finally, an effective kinetic model of ionic transport in nanopores is presented. We conclude with some final remarks.

\section{Methods}

The nanopore system setup considered in this work is shown in Fig.~\ref{fig1}(a) (see Fig.~\ref{fig1}(b) for its schematics). The system consists of a $\beta$-Si$_3$N$_4$ membrane immersed into a water solution of sodium and chlorine ions. The silicon nitride has a layered structure in the $z$ direction (along its $c$-axis), with two layers per unit cell. The corresponding lattice constant is $c=2.91$ \AA~\cite{wada81a}. In our calculations, the thickness of the  membrane was varied from $c$ to $8c$ ($8c$ is the typical thickness of the membranes used in studies of nano-ionic devices \cite{R5}). The radius of a cylindrical nanopore drilled through the membrane's center in most of our studies was $r = 6$~\AA (at this value of $r\sim r^*$ the nanopore is still in the Coulomb blockade regime while dehydration effects~\cite{Zwolak09a} are small~\cite{C2CP41641G,R17}) . Different radii and locations of the nanopore were also explored. An external electric field $E=V/d_a$ was applied along the $z$ direction, where $V$ is the bias voltage across the simulation cell region of length $d_a = 90$~\AA.

The ionic transport was studied by employing all-atom MD code NAMD2~\cite{R18}. The ion concentration, $m$, was varied up to 3 mol/L in order to clearly observe the ionic Coulomb blockade \cite{R16}. The periodic boundary conditions were employed in the directions perpendicular and parallel to the membrane. The CHARMM27 force field was used in our simulations. The Si$_3$N$_4$  atoms were confined in order to mimic the dielectric properties of Si$_3$N$_4$. A 1 fs time step was used. The temperature was kept at 275 K using a Langevin dampening parameter of 0.2 ps$^{-1}$ in the equations of motion. The van der Waals interactions were gradually cut off starting at 10 \AA~from the atom until reaching zero 12 \AA~away. The energy was initially minimized for 2 ps and then equilibrated for 100 ps at zero electric field. We subsequently ran our simulations for 15 ns in the NTV ensemble. The first 5 ns were used as a relaxation time to allow the system to reach the steady state (see Supporting information S1 for more details). The remaining 10 ns were used for analysis.

\section{Results and Discussion}

\subsection{Pore selectivity}

Similarly to previous MD investigations of ionic transport \cite{R16}, the nanopore surface in our study was not optimized and dangling bonds were not passivated \cite{Bermudez200511}. Consequently, the nanopore walls contain a number of immobilized free radicals having a strong influence on the ions flowing through the nanopore. The atomic structure in actual experiments is not known, but atomic surface corrugation with consequent local fields are expected.

We have observed that the nanopore selectivity changes with the nanopore diameter. Moreover, the nanopore position in the membrane (at fixed diameter) influences the current strength as well.
Fig. \ref{fig2} shows examples of the atomic structure of nanopore surface plotted for selected nanopore radii and positions. The surface of $r=5$~\AA~nanopore drilled at $x=0$  (Fig.~ \ref{fig2}(a)) contains lines of Si and N atoms in the $z$ direction such that the ratio of Si to N atoms is 4/8.
As the nanopore radius is enlarged, the atomic structure of its surface changes. Fig.~\ref{fig2}(b) shows that $r=6$~\AA~nanopore at $x=0$ (used in most of our runs) has 8/4 ratio of Si to N atoms. The shift of this nanopore to $x=6$~\AA~changes this atomic ratio to 4/9. The coordinate system with respect to the unit cell of Si$_3$N$_4$ is presented in Fig.~ \ref{fig2}(d).

Naively, one would expect that the nanopore selectivity is defined solely by the atomic composition of its surface (such as shown in Figs. \ref{fig2}(a)-(c)). In reality, however, the situation is more complex. In particular, for all locations of $r=6$~\AA~nanopore considered in this study ($x=0$, 2, 4, 6, 8, 10~\AA; $y=0$), the anionic selectivity was observed
with an insignificant contribution of cations to the current (see Fig.~\ref{fig2_2} inset). Moreover, it was found that the current strength depends on the nanopore location, as shown in Fig.~\ref{fig2_2}. For example, our calculations demonstrate that the current through the nanopore at $x=6$~\AA~is approximately two times smaller than that through $x=0$ nanopore.

We have also observed that the nanopore selectivity changes with diameter. For instance, we found that $r=5$~\AA~nanopore at $x=0$ is cation-selective (see Supporting information S2). Moreover, $r=10$~\AA~nanopore at $x=0$ is opened to both cations and anions with more anions contributing to the current (such larger nanopores also shows the Coulomb blockade). While the bipolar transport in larger diameter nanopores is expected, the transport in smaller nanopores is likely controlled by the charge of few layers of wall atoms. The calculations described in the next paragraph strongly support this statement (we observed a bipolar transport through $r=6$~\AA~nanopores when the charges of all membrane atoms were set to zero).


\subsection{Features of ionic transport}

When a bias voltage is applied across the system, the ions tend to drift in the direction defined by their polarity. In solution, the ions are covered by hydration layers that partially screen the electrostatic interactions among them. However, in order to enter a
sub-nanometer nanopore, the hydration layers need to be (partially) destroyed~\cite{Zwolak09a} as the nanopore geometry does not support atomic assemblies of the size larger than the nanopore diameter. Therefore, inside the nanopore, the role of ion-ion and ion-surface interactions becomes more pronounced.

In the case of the anion-selective nanopore from Fig.~\ref{fig2}(b) ($r=6$~\AA$\;$ and $x=0$), we have observed that the major contribution to the current is caused by chlorine ions (anions).  Taking a closer look at the ionic trajectories inside the nanopore, one can notice that the ions move through a set of discrete lattice states that can be linked to local minima of electrostatic potential due to membrane atoms. In Fig.~\ref{fig3}(a) we visualize this observation plotting the trajectories of chlorine ions inside the nanopore found at $m=1$ mol/L molarity. According to the right panel of Fig.~\ref{fig3}(a), the ions are stacked at discretized positions that are indicated by arrows.

The dynamical aspects of ionic transport through the nanopore involve the Brownian motion (stochastic jumps), ion-ion and ion-wall interactions, and drift in the applied field.
Some of these processes can be recognized in Fig.~ \ref{fig3}(a). In particular, the events (A) and (B) represent a simultaneous jump of two ions coupled through the Coulomb interaction. The events (C) and (D) exemplify single-ion stochastic jumps along and opposite to the stream, respectively.
In order to further explore the role of ion-ion and ion-wall interactions, we considered the case of an extremely narrow sub-nanometer nanopore of radius $r = 3.5$ \AA. It was observed that inside such a nanopore the ions stay in a crystalline arrangement thus forming an {\it ionic crystal-like structure}~\cite{ICC} (see Supporting Information S4 for more details).

We also note that, occasionally, sodium ions also enter the nanopore propagating in the opposite direction to the direction of chlorine ions (see Supporting Information S5 for additional details). These sodium ions give positive peaks to the distribution of charge density as shown in the right panel of Fig.~ \ref{fig3}(a). Figs. \ref{fig3}(c) and \ref{fig3}(d) show the distributions of ($x,y$) coordinates of chlorine and sodium ions when the ions are in the nanopore. The sodium and chlorine ions are strongly localized in proximity to N and Si atoms of the nanopore surface, respectively (see $r=6$~\AA$\;$ and $x=0$ case in Fig.~\ref{fig2}(b)).

\subsection{Current}

The saturated current-molarity relationship is a signature of the ionic Coulomb blockade~\cite{R16}.
 Fig.~\ref{fig4} presents the current calculated as $I = e_n/\tau$,
where $e_n$ is the net charge of ions transferred through the nanopore within the observation time interval $\tau$.
At smaller molarities, the current-molarity dependence is approximately linear.
At higher molarities, the current saturates and even decreases. Such current-molarity behavior is in agreement
with previous demonstrations of ionic Coulomb blockade~\cite{R16}.

In order to prove that the saturation is due to the trapping of ions inside the nanopore, we performed few simulations with uncharged membranes that did not reveal any signs of  current saturation (see Supporting information S3). In these simulations, the membrane thickness was varied from a length $c$ (2D membrane) to few $c$'s. For all values of $c$, the Coulomb blockade was restored as soon as the charges of membrane atoms were taken into account. Based on this observation, it seems reasonable to associate the Coulomb blockade with a potential barrier created by ions trapped in the nanopore repelling other incoming ions.
Fig.~\ref{fig4} also shows that MD simulation results becomes more scattered at higher concentrations. The increase in data scattering with concentration is most probably related to interplay of stochastic dynamics of trapped atoms with some other incoming ions.

In order to better understand the ionic transport in the nanopore, we investigated the effect of the membrane thickness on the ionic current.
For this purpose, we calculated the ionic current as a function of membrane thickness at two representative values of molarity of 1 and 2 mol/L (corresponding to the linear and Coulomb blockade regions in Fig.~\ref{fig4}, respectively) and a constant voltage
$V=0.5$ V applied across the system. The results of these calculations are presented in Fig.~\ref{fig5}.
It follows from Fig.~\ref{fig5} that in both cases the current decreases with an increase in the membrane thickness.

 Fig.~\ref{fig5} results can be well approximated by an effective circuit model of $L$ in-series connected resistors, where $L$ is the number of unit cells in the $z$ direction. The total current is then given by
\begin{equation}
I = \frac{V}{r_1+(L-1)R},
\label{eq8}
\end{equation}
where  $r_1$ is the resistance of the single-layer membrane that includes all the bulk and interfacial effects, and $R$ is the resistance of each additional membrane layer.
Fig.~\ref{fig5} demonstrates that Eq. (\ref{eq8}) curve (dashed line) describes quite well the MD simulation results. The best fit is obtained using the following parameter values:
$r_1 = $ 2.4 $\times 10^{8}$ $\Omega$ and $R =  $ 1.5 $\times 10^{8}$ $\Omega$ ( $m=1$ mol/L), $r_1 = $ 1.7 $\times 10^{8}$ $\Omega$ and $R =  $ 1.4 $\times 10^{8}$ $\Omega$ ($m=2$ mol/L). It is interesting that the value of $R$ is almost unaffected by the change in molarity. This indicates that the role of membrane thickness is in some sense ``universal'' and can be clearly distinguished from other effects. At the same time, the resistance $r_1$ is sensitive to the molarity and thus incorporates the effects of Coulomb blockade and access resistance~\cite{Hall75a}.

\subsection{Chain model of ionic transport}

There exist several theoretical models of ionic transport through nanopores such as the energy barrier~\cite{channelmodel1}, kinetic equation~\cite{channelmodel2}, and asymmetric association/dissociation~\cite{channelmodel3} models. In Ref. \cite{R16}, the Coulomb blockade was described by a capacitive barrier model.  Moreover, the effect of the selectivity filter charge was discussed in Ref. \cite{Kaufman15a}. However, to the best of our knowledge, all existing models of ionic Coulomb blockade do not take into account electrostatic interactions of transporting ions in sufficient details. Here, we develop a model that explicitly incorporates the ion-ion and ion-wall interactions. Our model replicates the ionic Coulomb blockade in narrow straight nanopores and captures the underlying dynamics of ionic transport.

According to our observations of ionic trajectories, the membrane atoms create a periodic potential for  mobile ions that propagate in a hopping-like fashion over the states corresponding to local potential minima. Therefore, we represent the nanopore as a lattice of sites and investigate the dynamics of ions on this lattice in the hopping transport regime.
Specifically, we consider a chain of $N$ sites (such as the one shown in Fig.~\ref{fig1}(c)) with a spacing $d_s= c/2$ corresponding to the period of Si$_3$N$_4$ layers. The chain is terminated by dummy sites that emulate the entrance and exit of the nanopore. The Cl ions are added to the dummy sites (from both ends of the chain) with an average constant rate $\gamma_{a}$. Two components of the ionic transport~\cite{R19} are considered: the temperature-induced stochastic jumps between neighboring sites and potential-induced drift (displacement), which are described by transition rates $\gamma_{b}$ and $\gamma_{d}$, respectively. Furthermore, it is assumed that $\gamma_{b}$ is constant while $\gamma_{d}$ is defined by several factors as described below.

\newcommand{\bvec}[1]{\mbox{\boldmath $#1$}}
Considering a small time interval $\Delta t$, the probability of jumping $P_{i \rightarrow i\pm1}$ from the site $i$ to site $i\pm 1$ is written as
\begin{equation}
P_{i \rightarrow i\pm1}=\left( \gamma_{b}+\gamma_{d}(i,i\pm 1) \right) (1-n(i\pm 1)) \Delta t,
\label{eq1}
\end{equation}
where $n(i\pm 1)=\{0,1 \}$ is the occupancy of the target site. Due to the factor $(1-n(i\pm 1))$, the transition is prohibited
if the target site is occupied. The probability of remaining on the same site is given by
\begin{equation}
P_{i \rightarrow i}=1-P_{i \rightarrow i+1} - P_{i \rightarrow i-1} .
\label{eq1_2}
\end{equation}

Next, the transition rate $\gamma_{d}(i,i\pm 1)$ is selected as a weighted difference of potentials at corresponding sites,
$\gamma_{d}=k( U_{i}-U_{i\pm 1})$, where $k$ is a constant. The  potential at site $i$, $U_i$, involves
contributions from the external electric field, nanopore surface atoms and ion-ion interactions, and is written as
\begin{equation}
U_i=Ez_i+U_{wall,i}+\kappa  \sum_{j\neq i}  \frac{e_j}{ \left| \bvec{R}_{j} - \bvec{R}_i  \right|}.
\end{equation}
Here, $z_i$ is the $z$ coordinate of the $i$-th site, $U_{wall,i}$ is the surface contribution to the potential, $\kappa =1/(4\pi\varepsilon)$, $\varepsilon$ is the effective permittivity in the nanopore, $e_j$ is the ionic charge at site $j$, $\bvec{R}_{j}$ and $\bvec{R}_{i}$ are the position vectors for sites $j$ and $i$. Moreover, we selected $U_{wall,i}=(-1)^iU_w$, where $U_w$ is a positive constant. Such form of $U_{wall,i}$
influences the probabilities to stay at even/odd sites simulating variations of nanopore wall potential. Our model calculations were based on the following set of parameter values: $N=16$ (selected based on the number of layers), $\gamma_{b}  = 9.05\times10^{2}$ ns$^{-1}$, $k = 3.1$ V$^{-1}$ns$^{-1}$, $U_{w} = 0.03$ V, $\Delta t=100$ fs, which were determined suitably to describe the dynamics of the transporting ions in Fig.~\ref{fig3}(a) and the current in Fig.~\ref{fig4}. The incoming ions rate $\gamma_a$ was selected in proportion to the molarity $m$ according to $m/(\gamma_{a}\Delta t) = 15$ mol/L. The value of $\gamma_{b}$ used in our simulations is lower than the typical bulk value because the diffusive behavior is reduced in the nanopore due to a strong ion-nanopore interaction ~\cite{Rdif1,Rdif2}.
In order to collect sufficient statistics, the system dynamics was modeled for $2.0\times 10^6$ steps.

Figure \ref{fig4} presents the current (blue solid line) calculated as the charge per unit time passing through the cross section of
the chain. One can observe the similarity between the results of MD
simulations and chain model. The simplified chain model also reproduces the relevant features of the transporting ions. Fig.~\ref{fig3}(b) shows ionic trajectories
computed using the chain model. In the right panel in Fig.~\ref{fig3}(b) we observe that the ions are accumulated with a near-uniform spacing, as pointed by arrows. This is also found in the MD simulation result in Fig.~\ref{fig3}(a). Moreover, as the results of time response in Fig.~\ref{fig3}(a), we see the forward transition generated by the additional ion (indicated by (A') and (B')) and the diffusive behavior (indicated by (C') and (D')). At lower concentrations, there are a few ions on the sites, and therefore, the Brownian motion mainly contributes to the ionic transitions. On the other hand, at higher concentrations, the sites are mostly occupied and strong ion-ion interactions play an important role in the transitions such as in the asymmetric association/dissociation model~\cite{channelmodel3}. The ions  trapped by $U_{wall,i}$ impede the flow of other ions resulting in the current saturation.

\section{Conclusion}

In conclusion, we have explored the ionic Coulomb blockade in nanometer-size nanopores, using all-atom MD simulations and a kinetic model. In our simulations, the ionic Coulomb blockade was observed in single and multi-layer membranes, for various nanopore sizes and locations. Importantly, we have unraveled the role played by surface effects in this regime, and have developed a simple kinetic model that captures the Coulomb blockade behavior observed in MD simulations. The ionic transport through the nanopore takes place in a hopping-like fashion
over a set of discretized states caused by local electric fields due to membrane atoms. For pores of small radii, this creates a slow-varying ``crystal-like'' structure of ions inside the pore, further reinforcing the Coulomb blockade behavior. Our results add to the understanding of ionic Coulomb blockade in realistic (sub-)nanometer size nanopores, and ionic transport in nanopores in general.

\section*{References}
\bibliography{literature}

\newpage

 \begin{figure}[h]
 \centering
 \includegraphics[width=0.9\textwidth]{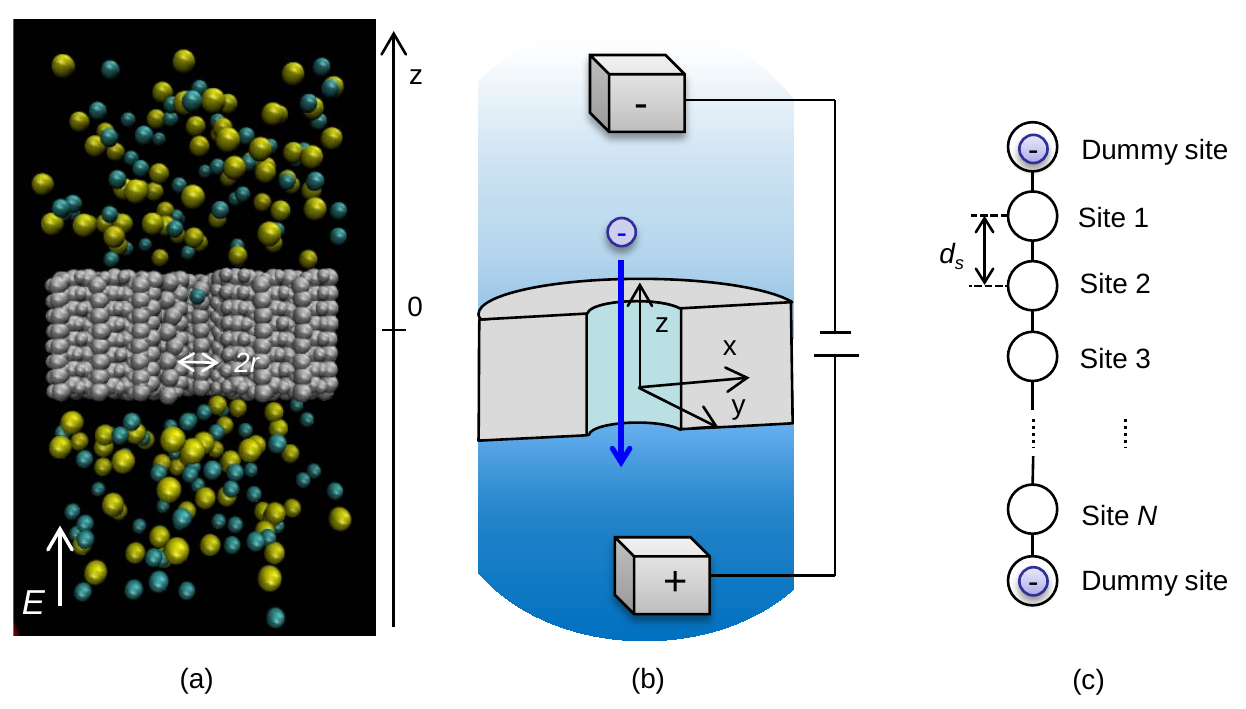}
 \caption{(a) Snapshot of a molecular dynamics simulation of 1 mol/L NaCl solution subject to an electric field $E$ created by an applied voltage of $V = 0.5$ V. A Si$_3$N$_4$ membrane that is indicated by white atoms has a nanopore with an opening of 6 \AA~radius. Chlorine and sodium ions are colored with aqua and yellow, respectively, and water molecules are invisible. (b) Schematic of the ionic transport through a nanopore. (c) Simplified chain model. Ions are added to the initial and last dummy sites with the in-flow ion rate $\gamma_{a}$. The site spacing $d_s$ along the $z$ direction is set at $c/2$ corresponding to the distance between Si$_3$N$_4$ layers. \label{fig1}}
 \end{figure}

 \newpage

 \begin{figure}[h]
 \centering \includegraphics[width=0.8\textwidth]{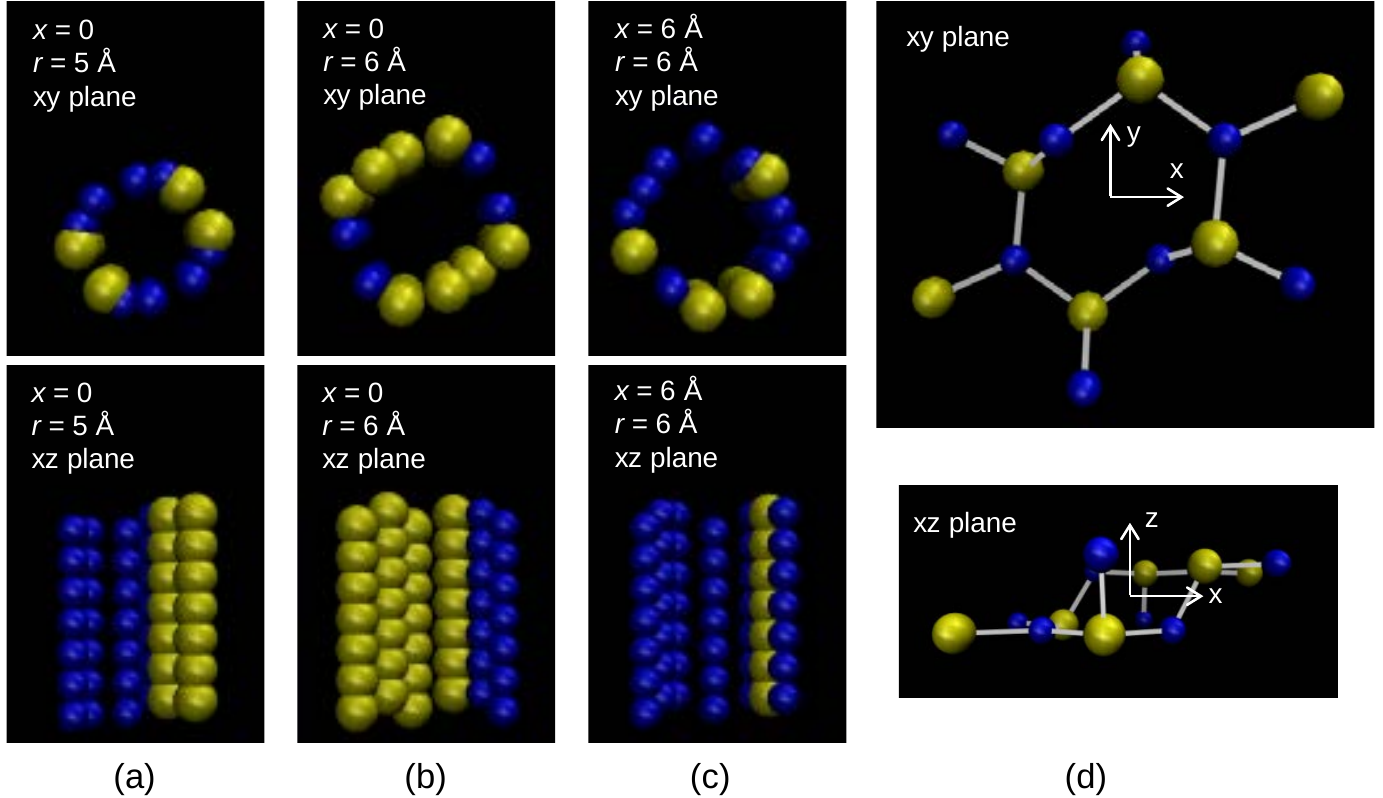}
 \caption{Atomic composition of the walls of different radii created at different locations: (a) $r=5$~\AA$\;$ and $x=0$, (b) $r=6$~\AA$\;$ and $x=0$, and (c) $r=6$~\AA$\;$ and $x=6$~\AA.  (d) Unit cell of Si$_3$N$_4$ around the coordinate origin. Si and N atoms are represented by yellow and blue spheres, respectively. \label{fig2}}
 \end{figure}

 \newpage

 \begin{figure}[h]
 \centering \includegraphics[width=0.45\textwidth]{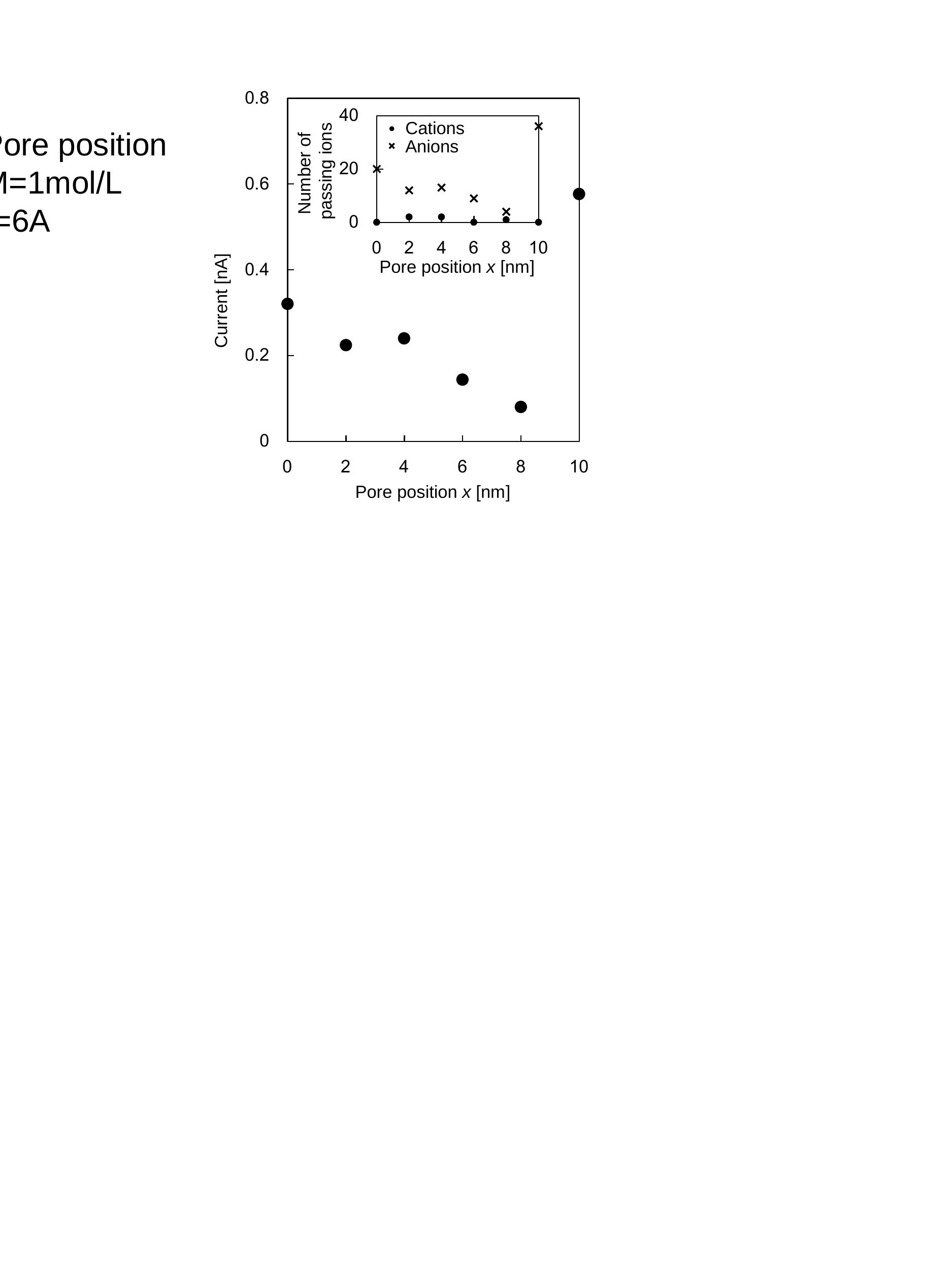}
 \caption{The current strength as a function of the nanopore location $x$ at $y=0$. This plot was obtained for the membrane thickness of $8c$ ($\sim 23$~\AA), the nanopore radius of $r=6$~\AA$\;$, the molarity of $m = 1$ mol/L and the  applied voltage of 0.5 V. The number of transporting ions through the nanopore is shown in the inset. The passing ions were counted for a time interval of 10 ns.} \label{fig2_2}
 \end{figure}

\newpage

 \begin{figure}[h]
 \centering \includegraphics[width=0.9\textwidth]{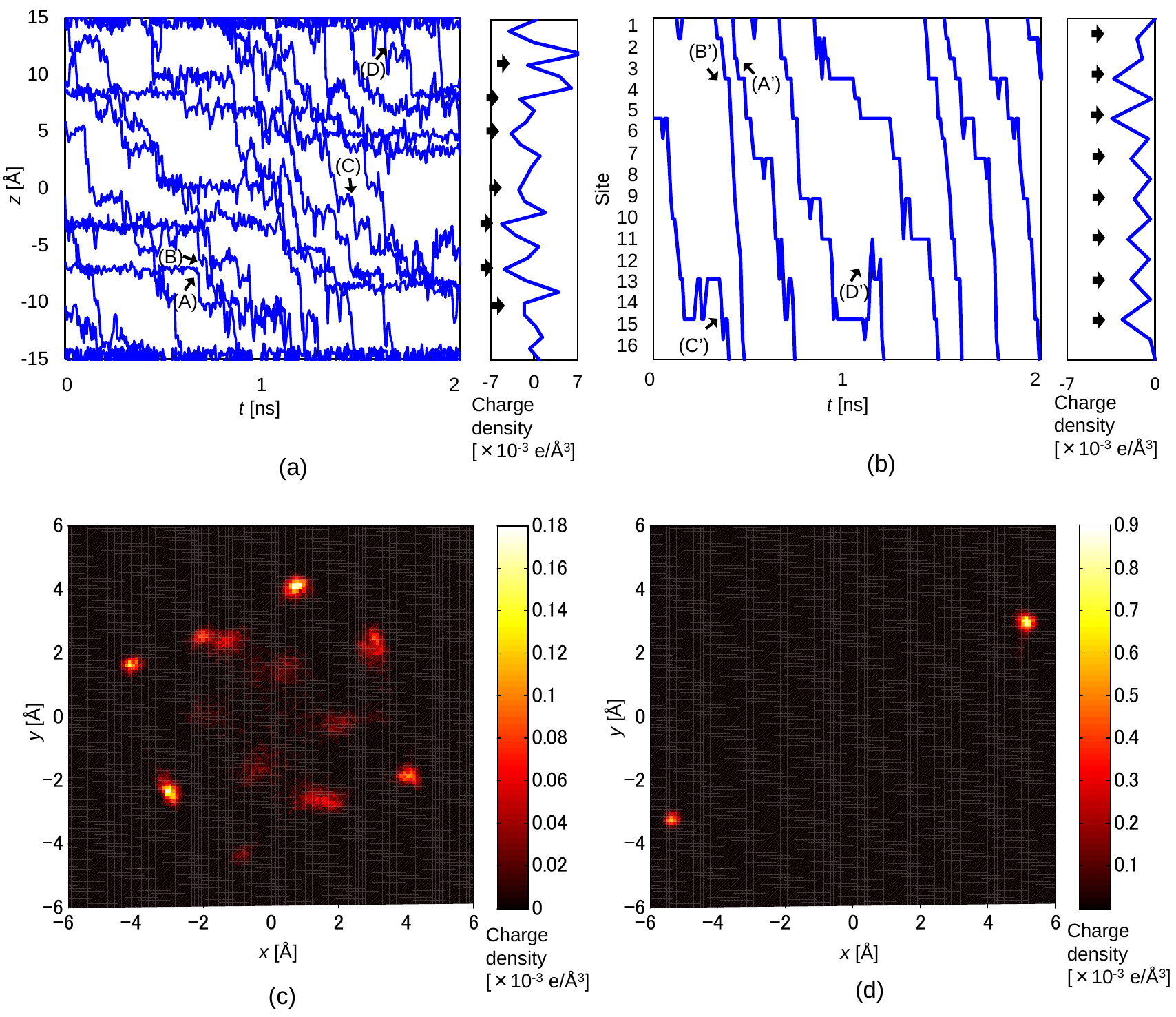}
 \caption{Trajectories of  chlorine ions inside the nanopore found in (a) the molecular dynamics simulation ($m = 1$ mol/L) and (b) chain model ($\gamma_{a}\Delta t = 0.25$), see text for details. The nanopore with $r=6$~\AA$\;$at $x=0$ is considered in the molecular dynamics simulation. The right panels show the charge density distributions for the time interval of the plotted trajectories. Distributions of ($x,y$) coordinates of (c) Cl and (d) Na ions inside the nanopore extracted from the molecular dynamics simulations. These plots were obtained for the membrane thickness of $8c$ ($\sim 23$~\AA)~ and the applied voltage of 0.5 V.  \label{fig3}}
 \end{figure}

 \newpage

 \begin{figure}[h]
 \centering \includegraphics[width=0.45\textwidth]{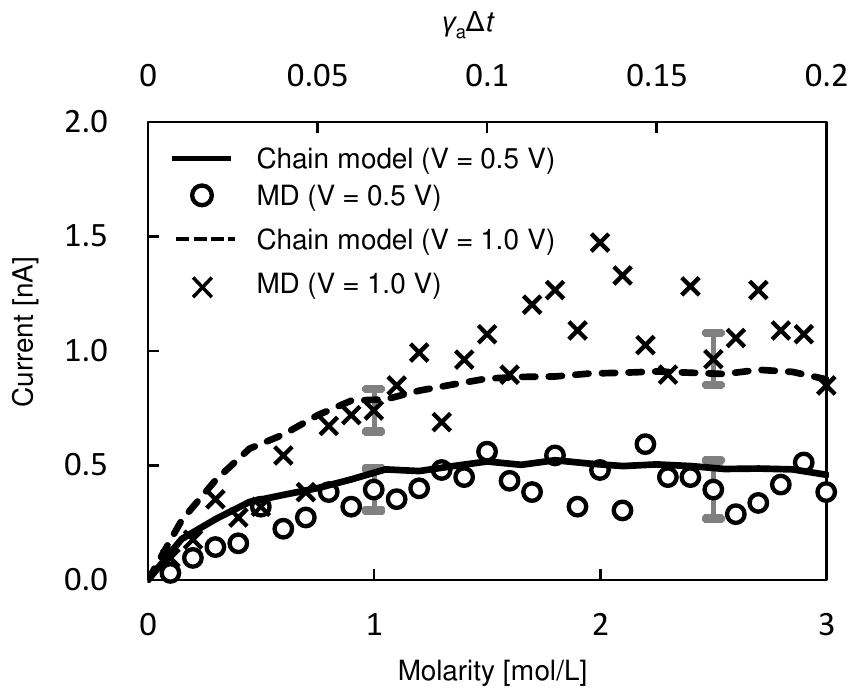}
 \caption{Comparison of currents obtained by molecular dynamics simulations (circles, bottom horizontal axis) and chain model (solid line, top horizontal axis). The nanopore with $r=6$~\AA$\;$at $x=0$ is considered in the molecular dynamics simulations. Thickness of the membrane is $8c$ ($\sim 23$~\AA)~ and the applied voltage is 0.5 V. The permittivity is assumed to be 6.4. We ran our MD simulations for 15 ns in the NTV ensemble, and ignore the initial 5 ns during which the system reaches the steady state. Moreover, 90\% confidence intervals are shown for 1 mol/L and 2.5 mol/L concentrations. \label{fig4}}
 \end{figure}

\newpage

 \begin{figure}[h]
 \centering \includegraphics[width=0.9\textwidth]{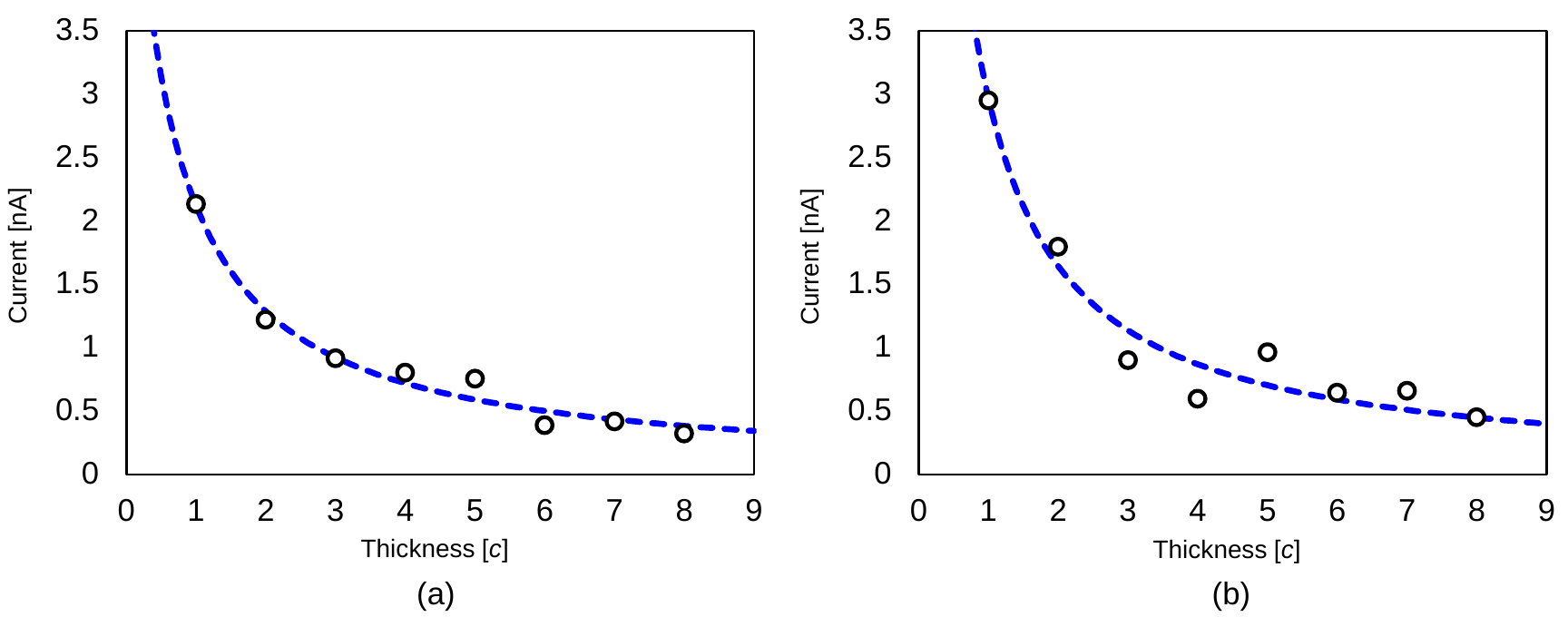}
 \caption{ Currents from molecular dynamics simulations (circles) and equivalent circuit model represented by Eq. (\ref{eq8}) (dashed line) for (a) $m=1$ mol/L and (b) $m=2$ mol/L. The nanopore with $r=6$~\AA$\;$at $x=0$ is considered in the molecular dynamics simulations. The applied voltage is 0.5 V. The membrane thickness is defined by the lattice constant of $c=2.91$ \AA. \label{fig5}}
 \end{figure}

\newpage

\noindent Supporting information

\setcounter{figure}{0}
\makeatletter
\renewcommand{\thefigure}{S\@arabic\c@figure}
\makeatother

\vspace{0.5cm}

\begin{center}

{\Large \bf Surface effects on ionic Coulomb blockade in nanometer-size pores}

\vspace{1cm}

 Hiroya Tanaka$^{1}$, Hideo Iizuka$^{1}$, Yuriy V. Pershin$^{2}$, and Massimiliano Di Ventra$^{3}$

\end{center}

\vspace{0.5cm}

\noindent $^{1}$ {\it Toyota Central Research \& Development Labs. Inc., Nagakute, Aichi 480 1192, Japan}

\noindent $^{2}$ {\it Department of Physics and Astronomy, University of South Carolina, Columbia, SC 29208, USA}

\noindent $^{3}$ {\it Department of Physics, University of California, San Diego, La Jolla, CA 92093, USA}

\vspace{0.5cm}

\noindent Email address: tanak@mosk.tytlabs.co.jp

\vspace{2cm}

\noindent Table of contents:

\vspace{0.3cm}

\noindent {\bf S1. Cumulative number of passing ions}

\noindent {\bf S2. Nanopore selectivity as a function of its radius}

\noindent {\bf S3. Ionic transport through uncharged membrane}

\noindent {\bf S4. Trajectories of chlorine ions in $r = 3.5$ \AA ~nanopore}

\noindent {\bf S5. Trajectories of sodium ions}

\newpage

\noindent {\bf S1. Cumulative number of passing ions}

\vspace{0.3cm}

\begin{figure}[h]
 \centering
  \includegraphics[width=0.9\textwidth]{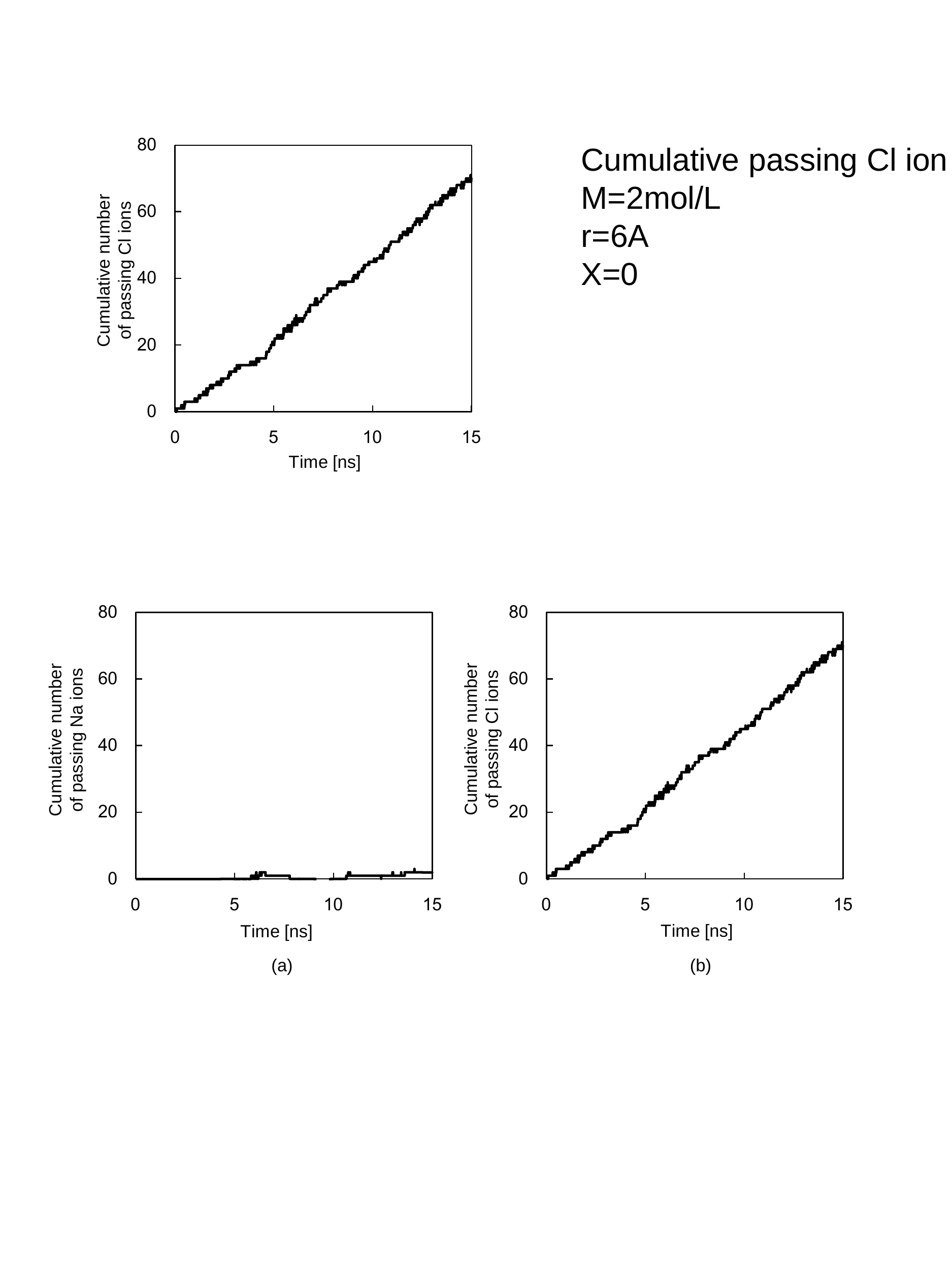}
 \caption{Cumulative number of the passing (a) Na, and (b) Cl ions obtained by the molecular dynamics simulation ($m=2$ mol/L). The nanopore with $r=6$~\AA$\;$at $x=0$ is considered. The thickness of the membrane is $8c$ ($\sim 23$~\AA)~ and the applied voltage is 0.5 V.  \label{figS1}}
 \end{figure}

Figure \ref{figS1} shows a cumulative number of the passing Na and Cl ions. The number of ions linearly increases with time. This indicates that about 5 ns, a time interval that was ignored in determining the current in our calculations, is enough in order to reach the steady state.

\newpage

\noindent {\bf S2. Nanopore selectivity as a function of its radius}

\vspace{0.3cm}

\begin{figure}[h]
 \centering
  \includegraphics[width=0.9\textwidth]{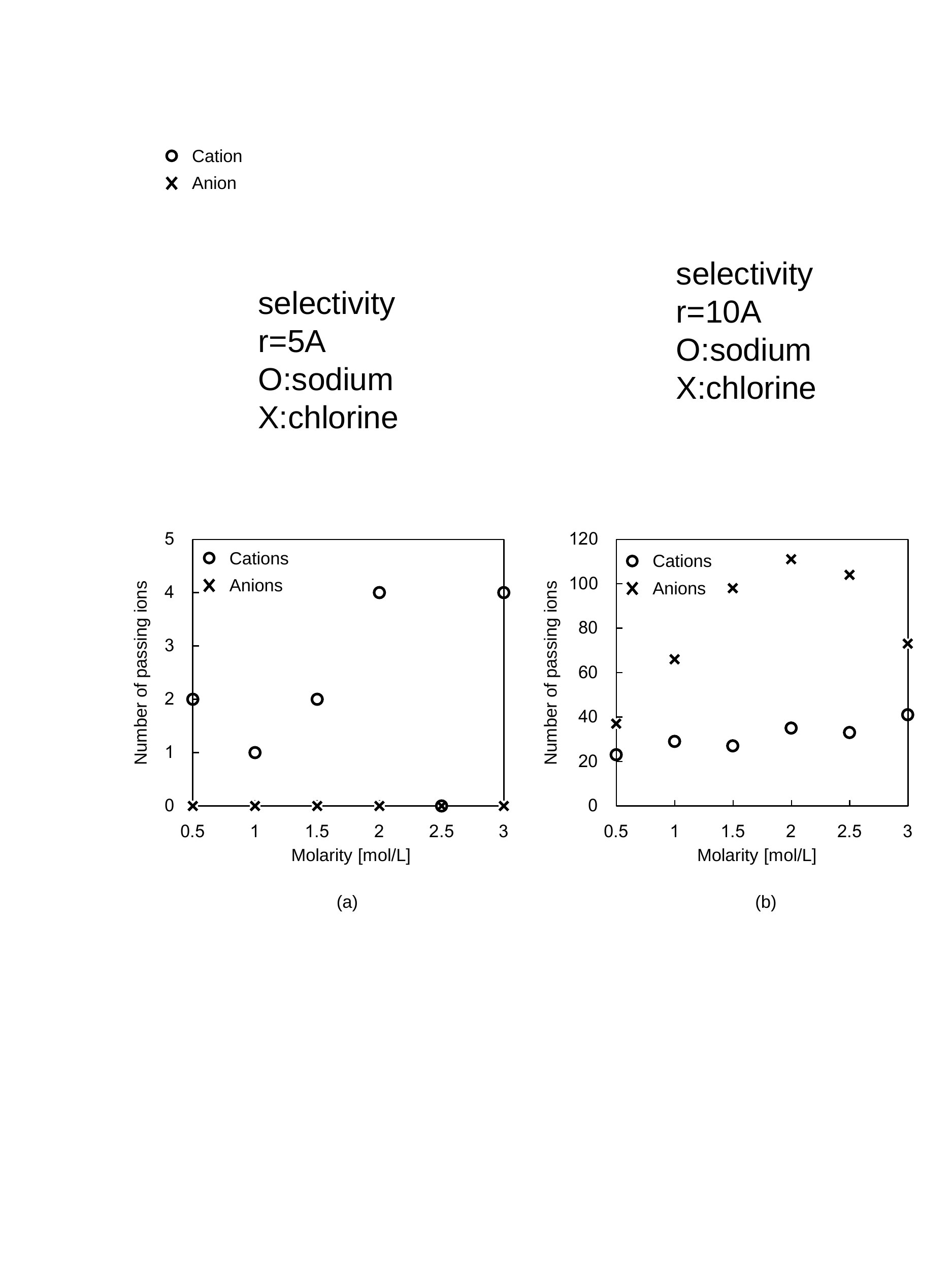}
 \caption{Number of transporting ions through the nanopore with (a) $r=5$~\AA, and (b) $r=10$~\AA, at $x=0$ for a time interval 10 ns. The molarity of the solution is 1 mol/L. These plots were obtained for the membrane thickness of $8c$ ($\sim 23$~\AA)~ and the applied voltage of 0.5 V. Numbers of ions were counted for the time interval of 10 ns after the system reached the steady state. \label{figS2}}
 \end{figure}

Figure \ref{figS2} shows a number of transporting ions through the nanopore. The cation selectivity is found in the the nanopore with $r=5$~\AA~at $x=0$ as shown in Fig.~ \ref{figS2}(a). On the other hand, in the case of the $r=5$~\AA~nanopore at $x=0$ shown in Fig.~\ref{figS2}(b), the transport of both cations and anions was observed, with more anions contributing to the current.

\newpage

\noindent {\bf S3. Ionic transport through uncharged membrane}

\vspace{0.3cm}

\begin{figure}[h]
 \centering
  \includegraphics[width=0.45\textwidth]{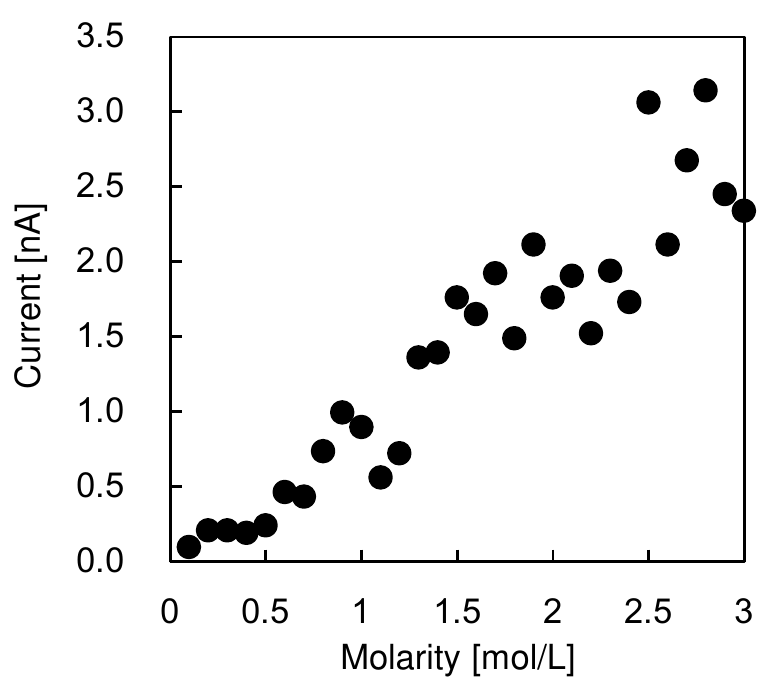}
 \caption{Ionic current through the nanopore ($r=6$~\AA$\;$, $x=0$) across an uncharged membrane.
 This plot was obtained for the membrane thickness of $8c$ ($\sim 23$~\AA)~ and the applied voltage of 5.0 V. \label{figS3}}
 \end{figure}

Figure \ref{figS3} presents the current calculated for membrane with zero charges assigned to its atoms. The current was calculated as $I = e_n/\tau$, where $e_n$ is the net charge of ions transferred through the nanopore within the observation time interval $\tau$. This indicates that the ionic Coulomb
blockade does not occur in the absence of membrane charges. In addition, the charges of the membrane atoms play an important role in the ionic Coulomb blockade.

\newpage

\noindent {\bf S4. Trajectories of chlorine ions in $r = 3.5$ \AA ~nanopore}

\vspace{0.3cm}

\begin{figure}[h]
 \centering
  \includegraphics[width=0.45\textwidth]{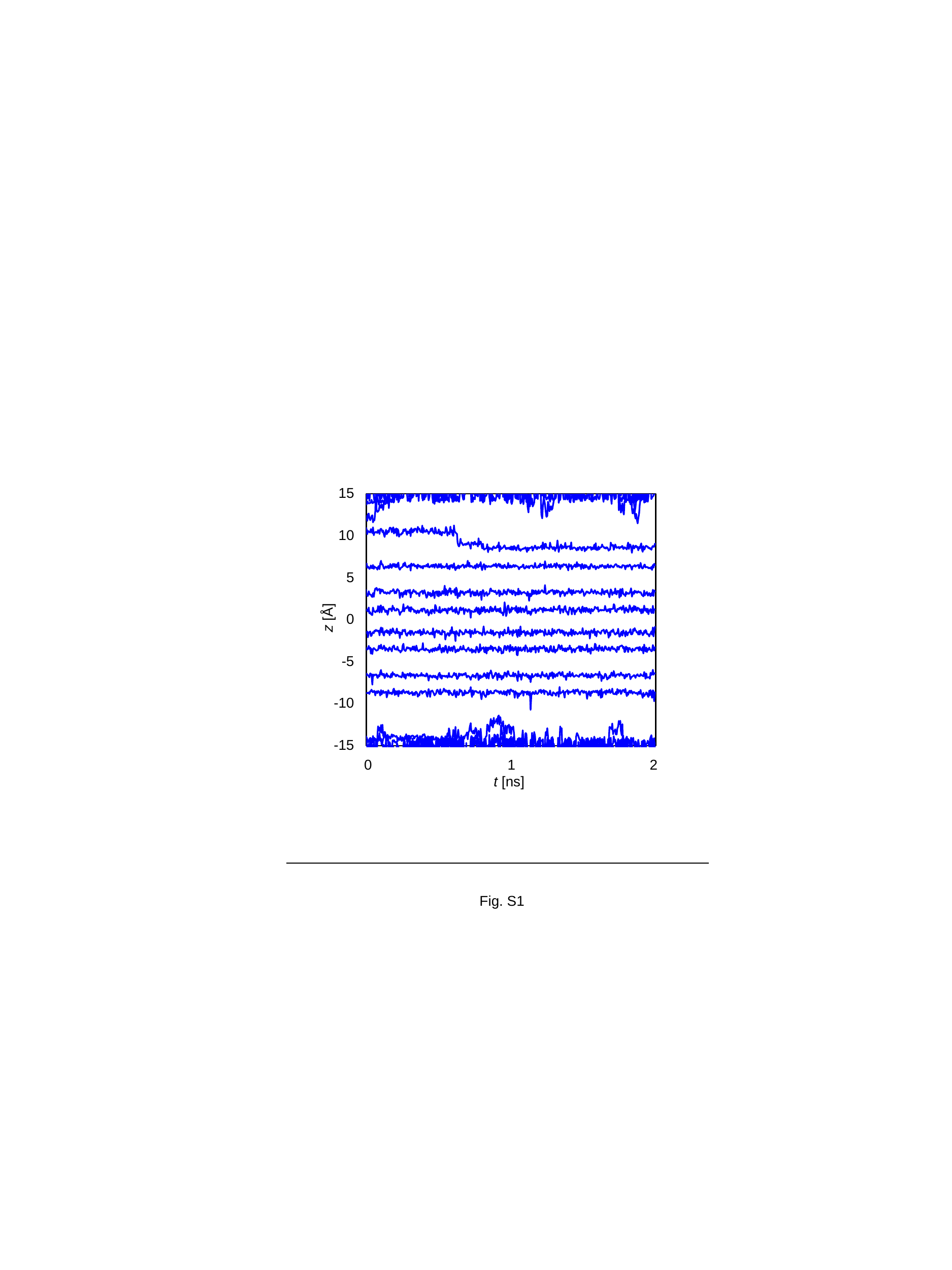}
 \caption{Trajectories of chlorine ions inside a nanopore of the radius of 3.5 \AA. This
  calculation was performed for the  membrane thickness of $8c$ ($\sim 23$~\AA), the molarity of 1 mol/L and the applied voltage of 0.5 V.  \label{figS4}}
 \end{figure}

Fig.~ \ref{figS4} presents the trajectories of chlorine ions found for the nanopore of the radius of 3.5 \AA~and the solution morality of  1 mol/L. The setup of this MD simulation corresponds to that presented in the caption of Fig.~\ref{fig1}. We observe that the ions form a well-defined arrangement due to strong Coulomb interactions in this extremely narrow nanopore. We note that the ionic current flow is crucially impeded by strong ion-ion interactions.

\newpage

\noindent {\bf S5. Trajectories of sodium ions}

\vspace{0.3cm}

\begin{figure}[h]
 \centering
  \includegraphics[width=0.45\textwidth]{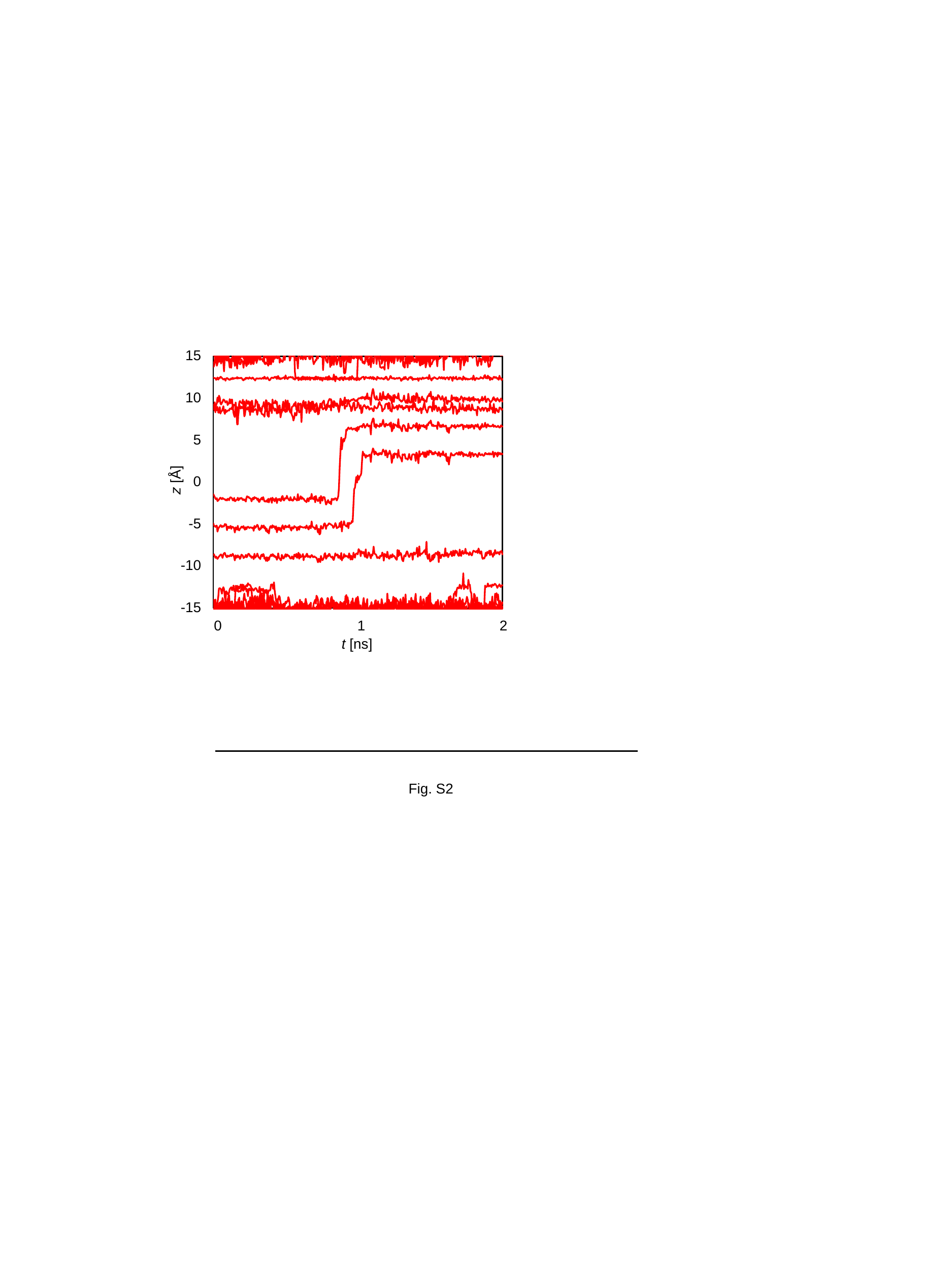}
 \caption{Trajectories of sodium ions. The data for this plot and Fig.~\ref{fig3}(a) were obtained in the same calculation. \label{figS5}}
 \end{figure}

Figure \ref{figS5} shows trajectories of sodium ions in the case when chlorine ions provide the main contribution to the current.
The sodium ions are responsible for the positive peaks in charge density distribution as shown in the right panel of Fig.~ \ref{fig3}(a).

\end{document}